\documentclass[10pt]{article}

\usepackage{graphicx}

\begin{document}

\title{AC and DC conductivity correlation: the coefficient of
Barton--Nakajima--Namikawa relation}

\author{{C. Tsonos\thanks{Corresponding author: tsonos@teilam.gr},
A. Kanapitsas, A. Kechriniotis and N. Petropoulos}}

\date{Department of Electronics, Technological Educational Institute 
 of Lamia, 3rd Km O.N.R. Lamia--Athens, 35100 Lamia, Hellas, Greece}

\maketitle

\begin{abstract}
It has been some time since an empirical relation,  which correlates  DC with  
AC conductivity and contains a loosely defined coefficient 
thought to be of order one, was introduced by Barton, Nakajima and 
Namikawa. In this work, we derived this relation assuming that the conductive 
response consists of a superposition of DC conductivity and an AC conductivity
term which materialized through a Havriliak--Negami dielectric function. The
coefficient was found to depend on the Havriliak--Negami shape parameters as 
well as on  the ratio of two characteristic time scales of ions motion which are
related to ionic polarization mechanism and the onset of AC conductivity. The results are
discussed in relation to other relevant publications and they also applied to a polymeric material. Both, 
theoretical predictions and experimental evaluations of  the BNN coefficient are in an excellent agreement,  
while this coefficient shows a gradual reduction as the temperature increases. 
\end{abstract}


\section{Introduction}
Nearly four decades ago, an empirical  relation was introduced by Barton,
Nakajima and Namikawa 
which is known as BNN relation \cite{Barton,Nakajima,Namikawa}. This expression
correlates the electrical conductivity to the dielectric strength of the lower 
frequency polarization mechanism through,  
\begin{equation}\label{bnn-1}
\sigma_0=p\varepsilon_0\Delta\varepsilon\omega_{max}~,
\end{equation}
where $p$ is a loosely defined parameter, expected to be of order 1, $\sigma_0$
is the apparent DC conductivity,  
$\Delta\varepsilon$ is the dielectric strength, $\omega_{max}$ is the angular
frequency which corresponds to the maximum value of dielectric losses and  $\varepsilon_0$ is
the permittivity of vacuum. The 
loss peak which associated to $\Delta\varepsilon$ is, in general, characterized
as broad. $\Delta\varepsilon$ may 
arise entirely from mobile charge effects
and not involve bulk dielectric effects at all. When Eq.~(\ref{bnn-1}) is
satisfied, both AC and DC conductivity may arise from the same type of charge transport mechanism
\cite{Lalanne,Furlani}. 

The BNN relation has played an important role in the analysis and the treatment
of frequency response dielectric data
\cite{Macdonald-1,Dyre,Porto,Dyre-Schroder,Sidebottom-1,Schroder-Dyre},
assuming that $ p \sim 1$ with most of the corresponding works dealing with the scaling
and universality issues of AC conductivity.  In the literature a large number of different 
conductive disordered materials have been found to satisfy Eq.~(\ref{bnn-1}).
The BNN relation is valid not only in amorphous solids, ionic glasses, single crystals and polymers but also in a
variety of other materials such as microporous systems \cite{Capaccioli} and proteins in hydrated
state \cite{Mijovic}.  The reported values of the coefficient $p$ vary
significantly, about three orders of magnitude from less than one up a few hundred
\cite{Barton,Nakajima,Namikawa,Lalanne,Furlani}. However, for a variety of materials,  the $p$ value 
falls mostly in the range from 0.5 to 10 \cite{Barsoukov-Macdonald}, 
while in the majority of the cases the $p$ value is near unity, as it is shown
as well in Fig. 3 of Ref. \cite{Dyre-Schroder}.

The  factors affecting the accurate estimation of $p$ value are discussed 
in Refs. \cite{Barsoukov-Macdonald, Macdonald-4}. However, some aspects should be
mentioned here which are related to the electrode effects. In the case of 
fully--blocking electrodes, if the data do not extend to the region where
$\sigma'(\omega)$ decreases towards zero in the lower--frequency plateau, the result might not define the
DC conductivity accurately. In the case of partial--blocking electrodes behavior
two regions of constant $\sigma'(\omega)$ values could possibly appear \cite{Krohns}. In
such cases it could be possible that the higher--frequency plateau region would 
lead to a more plausible $p$ estimate than the lower--frequency one, even though the 
latter is considered as the DC conductivity \cite{Krohns}. 

In the various models, which have been proposed for the description of the
dielectric response of disordered conductive materials, the BNN relation has been used as a testing
equation through the calculation 
of $p$ coefficient \cite{Dyre,Dyre-Schroder,Garcia,Macdonald-2}. The value of
$p$ is definitive, in order one to classify conductive materials. According 
to Hunt \cite{Hunt-1}, the role of Coulomb interactions in the derivation of the
BNN relation is of great importance, while the non--universality of the high frequency 
limit of the AC conductivity is incompatible  with a universality 
in the BNN relation \cite{Hunt-2}. However, various models proposed for a
particular kind of materials such as ionic glasses and 
disordered conductors, give universal values for the BNN coefficient $p$. Dyre
\cite{Dyre} obtains a value of $p=0.42$ by using the random energy barrier model. In a subsequent  
work Dyre and Schr{\o}der \cite{Dyre-Schroder} reported 
a value of $p=1.5 \pm 0.4$ for the simulation of the symmetric hopping model in
the extreme disorder limit. It has been pointed out by Macdonald \cite{Macdonald-1} that the
K1 conducting--system model could 
lead to a quantitative value for $p$, which depends on the value of $\beta_{1C}$
of the Kohlrauch--Williams--Watts (KWW) stretched 
exponential response function. For ion--conducting 
homogeneous glasses and single crystals with charge motion allowed in all three
dimensions, it has been shown that the only possible value is 
$\beta_{1C}=1/3$ and the resulting high frequency limiting response power law exponent is equal to 2/3
\cite{Macdonald-2}. According to these values, the BNN 
coefficient has a universal value of $p=1.65$, while in the framework of the K1 
model, as $\beta_{1C} \rightarrow 1$, $p$ should also approach unity in the
limit. In a recent paper Macdonald \cite{Macdonald-4} has provided a 
detailed analysis and $p$ estimates for the variety of conductive--system
models. These models involving a single fractional exponent, for an appreciable range of 
exponent values, show that the $p$ values are quite near 1. 

A modified BNN relation has been suggested by Dygas in \cite{Dygas}. It is
proposed that the values of modified $P$ coefficient are related to the spatial extent and time scale of
nonrandom local hopping of charge carriers. It also gave an expression of the BNN coefficient 
in the case of Cole--Cole dielectric behavior of ionic polarization mechanism.

The BNN equation quantifies the relation between short range and long range ions
motion of the AC response of conductive materials. To be specific the p coefficient reflects a measure of
the correlation between AC and DC conductivity. In the present work, we will attempt to derive the BNN
coefficient based on impedance spectroscopy formalism, as well 
as on widely used phenomenological and empirical relations and to discuss the results with relevant 
published works. The exact knowledge of the parameters on which 
the $p$ coefficient depends, is of great  importance, not only from the
fundamental point of view, but also for applications, 
because this could lead to design and development of a variety of materials with
predetermined dielectric and electrical properties.

\section{Theoretical considerations}

The complex conductivity $\sigma^*(\omega)=\sigma'(\omega)+j\sigma''(\omega)$ is
connected to the total  
complex dielectric constant
$\varepsilon^*(\omega)=\varepsilon'(\omega)-j\varepsilon''(\omega)$ via the
following relation
\begin{equation}\label{bnn-2}
\sigma^*(\omega)=j\omega\varepsilon_0\varepsilon^*(\omega)~.
\end{equation}              
In the above relation, if the contribution of the DC conductivity, $\sigma_0$,
is subtracted from 
$\varepsilon^*(\omega)$, then 
\begin{equation}\label{bnn-3}
\sigma^*(\omega)=\sigma_0 + j \omega  \varepsilon_0 \varepsilon_d^*(\omega)~,
\end{equation}       
where  $\varepsilon_d^*(\omega)$ represents the complex dielectric constant
caused only from the 
dynamic conductivity.

In ionic materials, the description of the real part of complex conductivity
spectra in the low frequency 
regime, below 100 MHz, and in the absence of electrode polarization effects, is
given by the  equation 
\cite{Dyre-etal,Sidebottom-2,Planes,Dutta,Elliott}
\begin{equation}\label{bnn-4}
\sigma'(\omega)=\sigma_0\left[1+\left(\frac{\omega}{\omega_0}\right)^n\right]~,
\end{equation}  
where $n$ (with $0<n<1$) is a constant. The characteristic frequency 
$\omega_0$ corresponds to the onset of AC 
conductivity. At this frequency, the real part of complex conductivity becomes
twice to that of the DC conductivity, $\sigma'(\omega_0)=2\sigma_0$. This last
equation has been 
introduced in Ref. \cite{Almond-West},
in order to describe crystals with defects and an 
activated number of charge carriers. Eq.~(\ref{bnn-4}) 
cannot be taken as a  model relation, because it is not able to reproduce the
individual AC 
response characteristics which are derived from other functions of impedance
spectroscopy (i.e. 
the peak and the dielectric strength, in $\varepsilon^*(\omega)$ formalism when
ionic dispersions 
take place). Eq.~(\ref{bnn-4}) is considered as a relation which  approximates
well 
only the frequency dependence of the real part $\sigma'$, since at low
frequencies, $\sigma'$ describes
the DC conductivity plateau of AC response, while  at high frequencies,
$\sigma'$ 
describes the well--known Jonscher power law behavior \cite{Jonscher}. In general, depending on 
the individual characteristics of
$\varepsilon_d''(\omega)$ responses, Eq.~(\ref{bnn-4})  
can or cannot  describe satisfactorily the function
$\sigma'(\omega)$ at the onset region.

If we assume that the real parts of 
Eqs.~(\ref{bnn-2}), (\ref{bnn-3}) and Eq.~(\ref{bnn-4}) are equal 
not only at the high frequency limit but also at 
$\omega=\omega_0$, then in these cases, from Eqs.~(\ref{bnn-2}), (\ref{bnn-3}) and  (\ref{bnn-4}) with 
$\omega=\omega_0$ we find 
\begin{equation}\label{bnn-5}
\sigma_0=\varepsilon_0\omega_0\varepsilon''(\omega_0)/2~~
\end{equation}
\textrm{and}~~
\begin{equation}\label{bnn-6}
\sigma_0=\varepsilon_0\omega_0\varepsilon_d''(\omega_0)~.
\end{equation}
So, in these cases the characteristic frequency $\omega_0$, should be defined also as the
frequency at which the losses from the 
DC conductivity are equal to the respective ones of the dynamic conductivity,
since 
$\varepsilon''(\omega_0)= 2\varepsilon_d''(\omega_0)$.

\section{The BNN relation}

In what follows, let us consider 
that, the conductive response of a disordered material in the frequency spectrum
under study, is characterized 
only by the existence of DC conductivity and an AC conductivity term. The latter
is considered that 
includes entirely contribution due to mobile ions effects. These effects should
lead to 
the appearance of a polarization mechanism in $\varepsilon^*$ formalism, which
should   
take place around the onset frequency $\omega_0$, with loss peak frequency,  
$\omega_{max}$, and strength $\Delta\varepsilon$. The real part of complex 
conductivity should  be given by using Eq.~(\ref{bnn-3}), as follows
\begin{equation}\label{bnn-7}
\sigma'(\omega) = \sigma_0 + \varepsilon_0\omega\varepsilon_d''(\omega)~.
\end{equation}            

The $\varepsilon_d''(\omega)$ should be considered as $\varepsilon''(\omega)$
conductive--system values, while  
for its description the well known and widely used Havrilak--Negami (H--N)
empirical dielectric function 
is used here as well \cite{Kremer},
\begin{eqnarray}\label{bnn-8}
\varepsilon_d''(\omega)=\frac{\Delta\varepsilon\sin(\beta\phi)}{\left[1+2(\omega
/\omega_{HN})^\alpha
\cos(\alpha\pi/2)+(\omega/\omega_{HN})^{2\alpha}\right]^{\beta/2}}&&\nonumber \\
&&
\end{eqnarray}                 
where
\begin{equation}\label{bnn-9}
\phi=\arctan\left(\frac{(\omega/\omega_{HN})^\alpha\sin(\alpha\pi/2)}{1+(\omega/
\omega_{HN})^\alpha
\cos(\alpha\pi/2)}\right)~.
\end{equation}                                       

The shape parameters take values in the range $0<\alpha,\beta \le 1$ and are
closely 
related to the slopes in 
$\log\varepsilon_d''$ vs. $\log\omega$ plots ($\varepsilon_d''\sim
\omega^\alpha$  at $\omega \ll \omega_{max}$
and $\varepsilon_d''\sim \omega^{-\alpha\beta}$  at $\omega \gg \omega_{max}$).
It should be mentioned here, that 
there exist models which have been proposed for the interpretation of the
limiting behavior of the H--N relaxation 
function, a behavior that is related to the slopes $\alpha$ and $-\alpha\beta$
\cite{Dissado, Schonhals}. The limiting 
case of $\alpha=\beta=1$ corresponds to Debye behavior with a single relaxation
time $\tau =1/\omega_{HN}$. The 
frequency $\omega_{HN}$ is related to  $\omega_{max}$ through 
\begin{equation}\label{bnn-10}
\omega_{max}=A\omega_{HN}~,
\end{equation}
where  
\begin{equation}\label{bnn-11}
A=\left(\frac{\sin(\alpha\pi/(2\beta+2))}{\sin(\alpha\beta\pi/(2\beta+2))}\right
)^{1/\alpha}~.
\end{equation} 

The total dielectric losses should be written as  
\begin{equation}\label{bnn-12}
\varepsilon''(\omega)=\varepsilon_d''(\omega)+\varepsilon_c''(\omega)~,
\end{equation} 
where $\varepsilon_c''(\omega)$ stands for the losses due to the DC 
conductivity, $\sigma_0$. So, from the real part of Eq.~(\ref{bnn-3}) we get 
$\sigma_0 = \varepsilon_0\omega\varepsilon_c''(\omega)$. At frequency $\omega=\omega_0$
and with the assumption that the real part of Eqs.~(\ref{bnn-3}) and (\ref{bnn-4})
are equal at $\omega = \omega_0$,   the relation $\varepsilon_c''(\omega_0)=\varepsilon_d''(\omega_0)$ holds 
(see Eq.~(\ref{bnn-6})). Therefore the relation 
$\sigma_0 = \varepsilon_0\omega_0\varepsilon_d''(\omega_0)$
will be valid and with the help of 
Eqs.~(\ref{bnn-8}) and (\ref{bnn-9}) we arrive at the BNN relation 
$\sigma_0=p \varepsilon_0 \Delta\varepsilon \omega_{max}$
where the BNN coefficient is equal to
\begin{eqnarray}\label{bnn-13}
p&=& \frac{\omega_0}{\omega_{HN}}\left(\frac{\sin(\alpha\beta\pi/(2\beta+2))}{\sin(\alpha\pi/(2\beta+2))}\right
)^{1/\alpha}\\ \nonumber
&&\times\frac{\sin(\beta\phi)}
{\left[1+2 (\omega_0/\omega_{HN})^\alpha \cos(\alpha\pi/2)+(\omega_0/\omega_{HN})^{2\alpha}\right]^{\beta/2}}~,
\end{eqnarray} 
where in this last expression  $\phi$ is given through Eq. ~(\ref{bnn-9})
at $\omega = \omega_0$, while in Cole--Davidson case ($\alpha=1$) takes 
its simplest form, $\phi=\arctan(\omega_0/\omega_{HN})$. 
 
. 
         
At the high frequency limit, $\omega \gg \omega_{HN}$,  Eq.~(\ref{bnn-8}) is given by   
$$\varepsilon_d''(\omega)\cong
\Delta\varepsilon\sin(\alpha\beta\pi/2)(\omega/\omega_{HN})^{-\alpha\beta}~,$$  
while Eq.~(\ref{bnn-9}) becomes 
$\phi=\alpha\pi/2$ and Eq. (\ref{bnn-7}) is written as
\begin{equation}\label{bnn-14}
\sigma'(\omega)\cong \sigma_0 +\varepsilon_0
\omega_{HN}\Delta\varepsilon\sin(\alpha\beta\pi/2)
(\omega/\omega_{HN})^{1-\alpha\beta}~.
\end{equation}       

The function $\sigma'(\omega)$ as given by Eqs.~(\ref{bnn-4}) and (\ref{bnn-14}), 
in the  high frequency limit of these two equations, $\omega \gg \omega_{HN}$,
should be the same. This means 
that the power law 
exponent of high frequency limit is $n=1-\alpha\beta$, while 
\begin{equation}\label{bnn-15}
\omega_0^n = \frac{\sigma_0}{\varepsilon_0\Delta\varepsilon
\sin(\alpha\beta\pi/2)}\omega_{HN}^{n-1}
\end{equation} 
holds as well. 
The previous relations  Eqs.~(\ref{bnn-14}) and (\ref{bnn-15}), coincide with the ones in 
Ref. \cite{Dygas}, where similar 
expressions have been derived previously. Then based on Eq.~(\ref{bnn-15})
and with the use of Eqs.~(\ref{bnn-1}), (\ref{bnn-10}) and (\ref{bnn-11}) 
the BNN coefficient should be expressed as
\begin{equation}\label{bnn-16}
p= \left(\frac{\omega_0}{\omega_{HN}}\right)^{1-\alpha\beta}\sin\left(\frac{\alpha\beta\pi}{2}\right)
\left(\frac{\sin(\alpha\beta\pi/(2\beta+2))}{\sin(\alpha\pi/(2\beta+2))}\right
)^{1/\alpha}~,
\end{equation} 
This last expression, in the Cole--Cole case, coincides with Eq.~(15) of \cite{Dygas}. Eq.~(\ref{bnn-16}) of 
this present work could be extracted as well from Eq.~(15) of Dygas work \cite{Dygas} with 
the help of our Eq.~(\ref{bnn-11}). However, it seems that an equation like our Eq.~(\ref{bnn-16}) 
has not been reported there, since the research  work in Ref. \cite{Dygas} is concentrated between 
other things in the physical meaning of 
their modified BNN coefficient.

\section{Discussion}

A distinction of the analysis presented above for the extraction of both
Eqs.~(\ref{bnn-13}) and (\ref{bnn-16}), in relation 
to other conductive--system dispersion models which are dealing with BNN
matters, is the use of the H--N dielectric function here. The H--N function should 
be considered as $\varepsilon''(\omega)$ conductive--system values. The fact,
that the H--N 
function is included in the AC conductivity term, is a consequence of its
definition. The coefficient 
of the BNN relation has been found to depend on the H--N 
parameters $\alpha, \beta$ and the frequency ratio $\omega_0/\omega_{HN}$ in both 
Eqs.~(\ref{bnn-13}) and (\ref{bnn-16}). The 
parameters $\alpha$ and $\beta$ represent the width and the skewness of the 
dielectric loss, $\varepsilon_d''(\omega)$, when viewed in a $\log(\omega)$ plot, and  
also describe the distribution function of the relaxation times \cite{Havriliak}. The 
ratio $\omega_0/\omega_{HN}$ should be viewed equivalently as the ratio of two
characteristic time lengths, $t_{HN}/t_0$, where $t_{HN}$ is the characteristic time length of 
the ionic polarization mechanism and $t_0$ is the respective one of the onset of
AC conductivity.

\begin{figure}
\includegraphics[height=8.3cm, width=8.7cm]{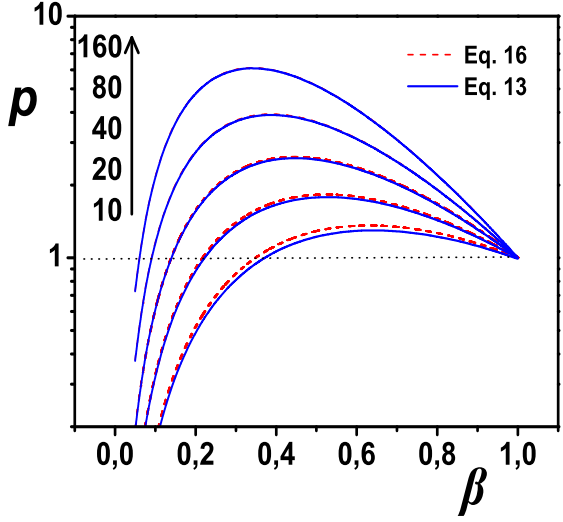}
\caption{\label{Fig1} The BNN coefficient $p$ as a
function of H--N parameter  $\beta$ for 
different frequency ratios $(10, 20,\dots, 160)$
according to Eqs~(\ref{bnn-13}), (\ref{bnn-16}). This figure represents the C--D
behavior.}
\end{figure}

\begin{figure}
\includegraphics[height=8.3cm, width=8.7cm]{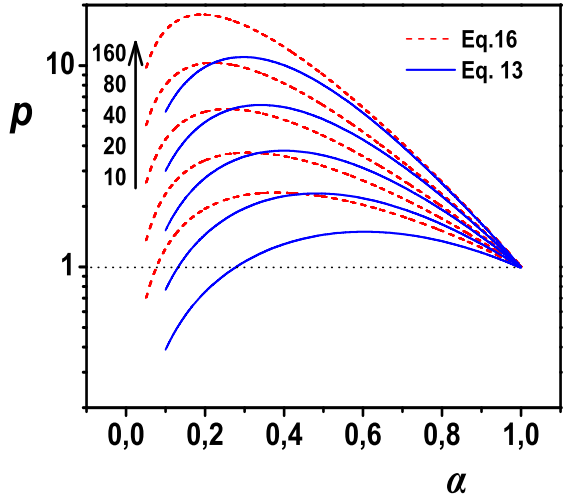}
\caption{\label{Fig2} The BNN coefficient $p$ as a function of H--N parameter $\alpha$
for different frequency ratios $(10,20,\dots, 160)$
according to Eqs~(\ref{bnn-13}), (\ref{bnn-16}). This figure represents the C--C
behavior.}
\end{figure}

In Figs.~ 1 and 2 the dependence of BNN coefficient $p$ versus the shape 
parameters $\alpha,\beta$ is given, for various values of the frequency ratio
$\omega_0/\omega_{HN}$ according to Eqs.~(\ref{bnn-13}) and (\ref{bnn-16}). Two 
characteristic cases could be observed there: the asymmetric Cole--Davidson
(C--D) behavior (Fig.1) and the symmetric Cole--Cole (C--C) one (Fig.2). As it is  
shown in both figures, for a behavior close to Debye one, the BNN coefficient
approaches 1. For particular values of the parameters $\alpha, \beta$ in both cases, the value of
$p$ increases by increasing the frequency ratio.

In both cases (C--D and C--C behavior) the value of $p$ increases up to a
maximum value as $\beta$ or $\alpha$ decrease, and then $p$ decreases as  $\beta$ or $\alpha$  
continues to decrease.  Values of $p<1$ are allowed in Eqs.~(\ref{bnn-13}) and (\ref{bnn-16}) for both behaviors.
We observe also that for the majority of $\beta$ or $\alpha$
parameter values, and for the  ratios  $\omega_0/\omega_{HN}$ shown in Figs. \ref{Fig1} and \ref{Fig2} the $p$ coefficient
takes values between 0.2 and  10.  As the ratio $\omega_0/\omega_{HN}$ gradually decreases, the $p$ coefficient tends to
take values near unity. 

As it can be seen in Fig.~\ref{Fig1}, the coincidence of the $p$ coefficient values ​​as 
specified by Eqs.~(\ref{bnn-13}) and (\ref{bnn-16}) is excellent. Two different functions 
have identical behavior in C--D behavior; only for lower frequencies ratio 
$\omega_0/\omega_{HN}$ is ​​there little difference in the values 
​​of $p$, and this difference is less than 0.07. It is also apparent in  Fig.~\ref{Fig2} in the C--C behavior,  
that the values ​​of $p$ coefficient as provided by Eqs.~(\ref{bnn-13}) and (\ref{bnn-16})  differ. But both $p$ 
values ​​are similar to the change as well the $\alpha$ parameter changes. The 
values ​​of $p$ coincide with each other to a greater range of high values ​​of 
$\alpha$ parameter, as the frequency ratio increases in C--C case.

To clarify the discrepancy in symmetrical C--C behavior, it is necessary to 
present a representative simulation. For parameter values $\alpha = 0.4$, $\beta = 1$ and 
frequency ratio $\omega_0/\omega_{HN}=40$, the values ​​of BNN coefficient according to 
Eqs.~(\ref{bnn-13}) and (\ref{bnn-16}) are found to be 3.78 and 5.38 respectively. Choosing the values 
​​$\sigma_0 = 10^{-6}$ S/m,  $\omega_{HN}=\omega_{max}= 100$ rad/s and the requirement 
to satisfy Eq.~(\ref{bnn-1}), the extracted values are: ​​$\Delta\varepsilon = 299$ from Eq.~(\ref{bnn-13})
and $\Delta\varepsilon = 210$ from Eq.~(\ref{bnn-16}).
\begin{figure}[h]
\includegraphics[height=8.3cm, width=8.7cm]{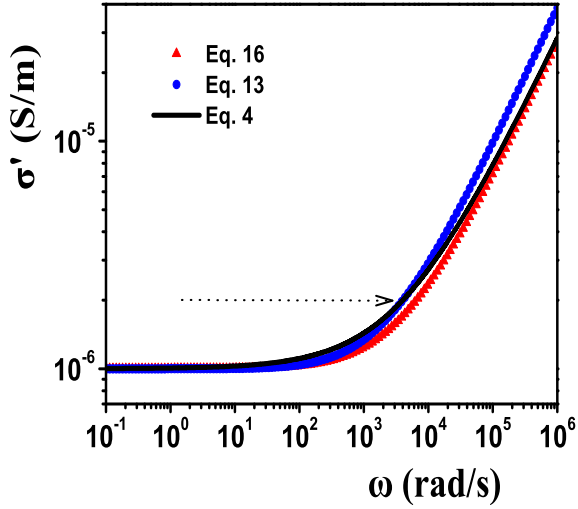}
\caption{\label{Fig3} Simulation curves of the  AC conductivity $\sigma'$ as a function of $\omega$ (details in the text).}
\end{figure}

The corresponding curves are presented in Fig.\ref{Fig3} 
with different ​​$\Delta\varepsilon$ values ​​of Eq.~(\ref{bnn-7}) (with Eqs.~(\ref{bnn-8}), (\ref{bnn-9})). Also 
presented the corresponding Eq.~(\ref{bnn-4}) with $\sigma_0=10^{-6}$ S/m, $\omega_0= 4000$ rad/s 
and $n = 0.6$ is also presented. It is clear that only the result of Eq.~(\ref{bnn-16}) produces a curve 
based on Eq.~(\ref{bnn-7})(with Eqs.~(\ref{bnn-8}) and (\ref{bnn-9})), which is described very well by the 
Eq.~(\ref{bnn-4}) at low and high frequencies limits. The resulting ​​$\Delta\varepsilon$ 
as calculated using  Eq.~(\ref{bnn-13}) fails to describe the power law frequency dependence of AC 
conductivity in the case of C--C dielectric behavior.

It should be noted here, that the additional assumption made in order to extract Eq.~(\ref{bnn-13})
is, that Eq.~(\ref{bnn-4}) and Eq.~(\ref{bnn-7}) (with Eqs.~(\ref{bnn-8}), (\ref{bnn-9})) should be 
equal (or at least approximate to a great extent), not 
only to high frequencies limit, but also at the characteristic frequency $\omega_0$. The failure 
of Eq.~(\ref{bnn-13}) in C--C dielectric behavior could possibly due to 
the fact that  the previous additional assumption cannot be applied in these symmetric responses.
As it is obvious in Fig.~\ref{Fig3} the power law frequency dependence of AC conductivity as given by 
Eq.~(\ref{bnn-4}) differs substantially at $\omega=\omega_0$, with the one given by Eq.~(\ref{bnn-7}) with the aid of 
Eq.~(\ref{bnn-16}). In contrast, Eq.~(\ref{bnn-7}),  based on the parameters 
of Eq.~(\ref{bnn-13}), although is not approximated well with power law (Eq.~(\ref{bnn-4}))
at the high frequencies limit seems to be approximated at frequency $\omega=\omega_0$  very 
well. The identification of $p$ values ​​in the 
asymmetric C--D behavior suggests that in these cases, 
the function $\sigma'(\omega)$  approximated well by Eq.~(\ref{bnn-4}) not only at the high 
frequency limit but also at frequency $\omega=\omega_0$. This should be 
the reason that the relations Eq.~(\ref{bnn-13}) and Eq.~(\ref{bnn-16}) although different,  actually coincide in the 
prediction of BNN coefficient, in C--D behavior.

\begin{figure}[h]
\includegraphics[height=8.4 cm, width=8.7cm]{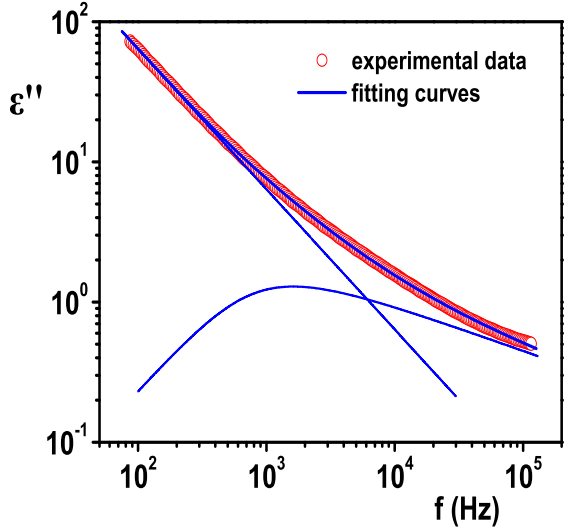}
\caption{\label{Fig4}  The imaginary part $\varepsilon''$ of the complex dielectric constant  as a function of  
of frequency $f$ at a temperature 363 K for a crosslinked polyurethane.  
In the figure is shown the total fitting as well as  the contribution of the polarization mechanism  
and the DC conductivity according to fitting process are shown.}
\end{figure}

In what follows trying to achieve contact with real materials, we will refer to a case study of dielectric response of a 
crosslinked polyurethane (PUR) which satisfies the BNN relation.
The $\varepsilon''(\omega)$ experimental data of PUR \cite{Kanapitsas}
are fitted with a sum of H--N expression (Eqs.~(\ref{bnn-8}) and (\ref{bnn-9})) and the 
term $A\omega^{-k}$  which represents the contribution of the DC conductivity, 
$\sigma_0$, to the dielectric losses spectra with value of $k = 1$. 
A representative fit is shown  at a temperature 363 K in Fig.~\ref{Fig4}. 
The dielectric dispersion exhibits, not only at this temperature, C--D behavior. 
The values ​​of $\sigma_0$ where resolved from the extrapolation of low 
frequencies $\sigma'(\omega)$ plateau at 1 mHz. The $f_0$  values ​​($\omega_0 = 2\pi f_0$) were 
taking these corresponding to values ​​$2\sigma_0$ according to $\sigma'(\omega)$ data. It 
should be noted that the dipolar dispersions appear at higher enough 
frequencies \cite{Kanapitsas} and so their contribution to $\varepsilon''(\omega)$  and $\sigma'(\omega)$ 
frequency spectrum studied here is negligible. Also during  
fitting process of $\varepsilon''(\omega)$ data, as well as the estimation of the DC 
conductivity, $\sigma_0$, we have been restricted  at a suitable low frequencies region, where the influence of the 
electrode effects could be considered as  negligible. The parameter values resulting 
from the fitting process in $\varepsilon''(\omega)$ data, as well as the 
values of $\sigma_0$ and $f_0$ which have been calculated from $\sigma'(\omega)$ data are given in Table~\ref{Table1}.
\begin{table}[h]
\caption{The parameter values of  $\alpha$, $\beta$, $\Delta\varepsilon$ and $f_{HN}$ resulting
from the fitting process in $\varepsilon''(\omega)$ data as well as the 
values of $\sigma_0$ and $f_0$ for crosslinked polyurethane at different temperatures $T$. 
We consider that the values of $\sigma_0$ and $f_0$  have been estimated 
with high accuracy and therefore those are given without errors.}
\centering
\begin{tabular}{c c c c c c c c}
\hline\hline
$T$(K) &  $\alpha$ & $\beta$ & $\Delta\varepsilon$ & $f_{HN}$ (Hz)& $f_{max}$ (Hz) & $f_0$ (Hz)& $\sigma_0$ (S / m)\\ 
[0.5ex]
\hline
323 & $1\pm 0.00$ & $0.39 \pm 0.01$ & $4.9 \pm 0.1$ & $62 \pm 5$ & $132 \pm 11$ & 560 & $3.50\times 10^{-8}$ \\
333 & $1\pm 0.00$ & $0.37 \pm 0.01$ & $4.8 \pm 0.1$ & $123 \pm 5$ & $272 \pm 11$ &1185 & $7.00 \times 10^{-8}$ \\
343 & $1\pm 0.00$ & $0.34 \pm 0.01$ & $4.9 \pm 0.1$ & $198 \pm 7$ & $470 \pm 17$ & 2178 & $1.27 \times 10^{-7}$\\
353 & $1\pm 0.00$ & $0.33 \pm 0.01$ & $4.8 \pm 0.1$ & $367 \pm 11$ & $894 \pm 27$ & 3560 & $2.15 \times 10^{-7}$\\
363 & $1\pm 0.00$ & $0.33 \pm 0.01$ & $4.7 \pm 0.1$ & $656 \pm 20$ & $1597 \pm 48$ & 5501 & $3.48 \times 10^{-7}$\\ 
\hline
\end{tabular}
\label{Table1}
\end{table}

\begin{figure}[h]
\includegraphics[height=8.3cm, width=8.7cm]{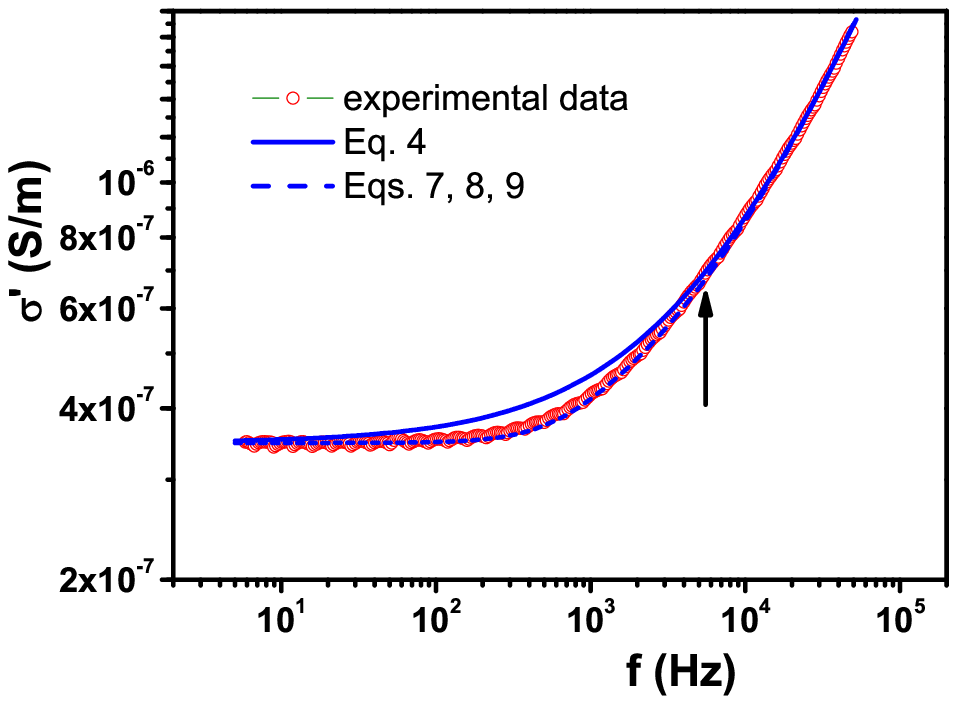}
\caption{\label{Fig5} The AC conductivity $\sigma'$ as a function of frequency $f$ at 363 K 
for a crosslinked polyurethane. The lines  correspond to simulation curves according 
Eq.~(\ref{bnn-4}) and Eqs.~(\ref{bnn-7}), (\ref{bnn-8}), (\ref{bnn-9}), and using the 
corresponding values of Table~\ref{Table1}.}
\end{figure}

Experimental data of $\sigma'(\omega)$ are presented in Fig.~\ref{Fig5}. In the same figure, the 
simulation curves of Eq.~(\ref{bnn-4}) and Eq.~(\ref{bnn-7}) 
(with Eqs.~(\ref{bnn-8}), (\ref{bnn-9})) are shown according to 
the parameters values ​​of Table~\ref{Table1}. Despite the fact that Eq.~(\ref{bnn-4}) does 
not  describe adequately the behavior of the rest two curves near the 
onset region of AC conductivity, there is a very good match in all 
curves for frequencies $\omega \ge \omega_0$ .

The values ​​of the coefficient $p$, as 
calculated based on the predictions of Eq.~(\ref{bnn-13}) and Eq.~(\ref{bnn-16}) as well as on 
the BNN relation, Eq.~(\ref{bnn-1}), are presented in Table~\ref{Table2}.  
In order to do these calculations the corresponding values contained ​in  Table~\ref{Table1} have been used. 
Although the temperature window is relatively small , it is 
evident that $p$ coefficient shows a gradual reduction, with increasing 
temperature, in all calculations included in Table~\ref{Table2}. The change of $p$, 
throughout the temperature range, appears to be 0.15--0.19.
At each temperature the  $p$ values as they have been calculated using the three equations, 
show an excellent agreement at  all temperatures
\begin{table}[h]
\caption{The values of $p$ coefficient at various temperatures.}
\centering
\begin{tabular}{c c c c  }
\hline\hline
$T$(K) & $p$ (Eq.~(\ref{bnn-16}))& $p$ (Eq.~(\ref{bnn-13})) & $p$ (Eq.~(\ref{bnn-1}))\\ [0.5ex]
\hline
323 & $1.04\pm0.05$ & $0.97\pm0.06$ & $0.98\pm0.07$ \\
333 & $1.03\pm0.03$ & $0.97\pm0.03$ & $0.97\pm0.04$ \\
343 & $1.05\pm0.03$ & $0.99\pm0.03$ & $0.99\pm0.03$ \\
353 & $0.93\pm0.02$ & $0.88\pm0.02$ & $0.90\pm0.03$ \\
363 & $0.85\pm0.02$ & $0.79\pm0.02$ & $0.83\pm0.03$ \\[0.5ex] 
\hline
\end{tabular}
\label{Table2}
\end{table}
According to this analysis, in order for the coefficient $p$ to be temperature independent, 
the parameters $\alpha$, $\beta$ and $\omega_0/ \omega_{HN}$ must be temperature independent, 
or in the case that they are temperature dependent, their contributions on $p$ value 
must be such that they cancel each other. In any other case the coefficient 
of $p$ is expected to be temperature--dependent. 

In Ref. \cite{Dygas},  experimental data have been mentioned, which are relevant to Havriliak--Negami 
dielectric function and the characteristic frequency ratio in three different systems which satisfy the BNN 
relation. The BNN coefficient is calculated according to original BNN relation, and it has  been found 
that  it is equal to 7.1, 1.34 and 2.1 for a single crystal, a type of glass and a PEO polymer 
electrolyte respectively (for details see Table 1 of Ref. \cite{Dygas}). Based on the values of 
Table 1 of Ref. \cite{Dygas}, the BNN coefficient was found to be 4.6, 1.15 and  1.7 by using  
Eq.~(\ref{bnn-13}),   while by using Eq.~(\ref{bnn-16}) it was found to be 4.8, 1.32 and  2.1 for a single 
crystal, glass  and PEO polymer electrolyte respectively. We observe that the values 
found using  Eq.~(\ref{bnn-16}) are identical to those based  
on the original BNN relation for glass and PEO polymer electrolyte, while those of Eq.~(\ref{bnn-13})
result in values close enough as well. However, in 
the case of the single crystal, although Eq.~(\ref{bnn-13}) and Eq.~(\ref{bnn-16})
result in  very close values, these values differ significantly from the one  extracted 
using the original BNN relation. This  difference is possibly due 
to uncertainties during the fitting process. The fitting of Havriliak--Negami function
in $\varepsilon''(\omega)$ and $\varepsilon'(\omega)$
data after subtraction of the DC conductivity and electrode effects
respectively,  lead generally in scattering of the data points at
frequencies lower than the ones which correspond to the peak of the dielectric
dispersions. This fitting process has as a result 
uncertainties in the estimation of the Havriliak--Negami parameters. 

As can be seen observing the values ​​of Table~\ref{Table1}, 
the relation in Eq.~(\ref{bnn-15}) is satisfied with sufficient accuracy. Note here that the strength of 
this relationship requires the power law Eq.~(\ref{bnn-4}) to approach very well the function 
$\sigma'(\omega)$ only at the higher frequencies limit, and thus its validity is more general . From 
Eq.~(\ref{bnn-13}) and Eq.~(\ref{bnn-16}) is evident that there are many possible $p$ values. Also, for a 
given value of BNN coefficient, there are several possible combinations of 
parameters $\alpha$, $\beta$,  and $\omega_0/ \omega_{HN}$, according to 
Eqs.~(\ref{bnn-13}) and (\ref{bnn-16}).  In these cases, 
not only the high frequencies slope of function $\sigma'(\omega)$, but also 
the ratio $\sigma_0/\omega_0 \Delta\varepsilon$ could be different with consequences in 
scaling diagrams of AC conductivity. Of course one question 
is if all these different combinations of parameters may correspond
to the response of the real materials. Assume for simplicity that the ionic 
polarization mechanisms of a category of materials, characterized by 
particular values $\alpha$, $\beta$ and $\omega_0/ \omega_{HN}$. Then, these materials 
characterized by a single value of $p$. This seems to be compatible  with 
the universal value of $p$, as proposed 
by Macdonald \cite{Macdonald-2}  for ion--conducting glasses and single
crystals. His conclusion is based on the observation that these materials are characterized by the same
shape parameter $\beta_{1C}$ of KWW response function and also the same 
characteristic frequencies ratio of K1 model. This indicates that H--N and 
KWW shape parameters are closely related, as well as the characteristic
frequencies ratio of our approach ($\omega_0/\omega_{HN}$) to that of the 
K1 model. Particular values ​​of the parameters $\alpha$, $\beta$ 
and $\omega_0/ \omega_{HN}$ implies 
that the corresponding materials not only have same $p$ value but also 
have the same ratio $\sigma_0/\omega_0 \Delta\varepsilon$, according to Eq.~(\ref{bnn-15}). Of course 
there are many possible combinations of $\sigma_0$, $\omega_0$ and  $\Delta\varepsilon$ values ​​which 
keep this ratio constant. But only some of these combinations 
correspond to the response of real materials.

In a recent paper Dyre {\textit{et al.} \cite{Dyre-etal} applied  the
fluctuation--dissipation 
theorem in ionic conductors. According to their calculations, the DC conductivity is  
\begin{equation}\label{bnn-17}
\sigma_0=\frac{n (t_0)q^2}{6k_BT}\frac{\langle\Delta r^2(t_0)\rangle}{\gamma
H}\omega_0~,
\end{equation}                   
where $q$ denotes the ions charge, $k_B$ is the Boltzmann constant, $T$ is 
the absolute temperature, $n(t_0)$ is the number density of mobile ions at 
time $t_0 =1/\omega_0$ ($\omega_0$ being  the onset frequency),  $\langle\Delta
r^2(t_0)\rangle$ is 
the single-particle mean square displacement, while  $\gamma$ and $H$ are
numbers which are roughly of order one. 
Roling {\textit{et al.} \cite{Roling} using linear response theory have found 
a relation between the ionic dielectric strength, $\Delta\varepsilon$, and the
mean--square displacement of the center of the mobile ions,
$\langle\tilde{R}^2(\infty)\rangle$, in ionic glasses conductors
\begin{equation}\label{bnn-18}
\varepsilon_0 \Delta\varepsilon=\frac{N_V
q^2}{6k_BT}\langle\tilde{R}^2(\infty)\rangle~,
\end{equation}                               
where $N_V$ is referred as the number density of mobile ions. Similar
expressions to that of Eq.~(\ref{bnn-18})
have already been reported in the literature \cite{Sidebottom-2,Macdonald-3}.
Now, substituting the quantity 
$\omega_0/\sigma_0$  as given by Eq.~(\ref{bnn-17}) and that of
$\varepsilon_0\Delta\varepsilon$ as given in 
Eq.~(\ref{bnn-18}), back into Eq.~(\ref{bnn-15}),  we arrive at the following
expression
\begin{equation}\label{bnn-19}
\frac{\omega_0}{\omega_{HN}}=\frac{t_{HN}}{t_0}=\left[\gamma H
\sin(\frac{\alpha\beta\pi}{2})
\frac{N_V}{n(t_0)}\frac{\langle\tilde{R}^2(\infty)\rangle}{\langle\Delta
r^2(t_0)\rangle}\right]^{\frac{1}{\alpha\beta}}~.
\end{equation} 
The factor $\gamma H\sin(\alpha\beta\pi/2)$ in the previous equation takes
values close to unity, for 
typical broad ionic dispersions. The characteristic frequencies or times ratio
depends on the number densities 
of mobile ions at different time scales which characterize
the AC response, as well as to the corresponding distance lengths of ions
motion. 
The other factor affecting the characteristic ratio is the product
$\alpha\beta$ 
which represents the high frequency absolute slope of ionic dispersion in 
$\log\varepsilon_d''$ vs. $\log\omega$ plot, and it is connected to the high
frequency limiting response power law exponent of $\sigma'$~. The lower the value of $\alpha\beta$,
the stronger its influence on the characteristic ratio is. As a consequence of this,different values of the 
power law exponent in the high frequency limit of AC conductivity, lead to
different values of BNN coefficient, in accordance with the conclusions about the
non--universality of $p$ value, as has been suggested by Hunt \cite{Hunt-2}. From Eq.~(\ref{bnn-19}), which is based on two 
different models \cite{Dyre-etal, Roling}, it is obvious that particular values 
​​of $\alpha$, $\beta$ and $\omega_0/ \omega_{HN}$ correspond to a specific correlation of the number densities 
as well as of the distance length of mobile ions at different time scales, which 
affect the macroscopic characteristic of long and short range ions motion.

Finally, it should be noted
that the high frequency slope of $\sigma'$, is $n=1-\beta_{KWW}$, where $\beta_{KWW}$ is the 
corresponding KWW stretched exponential parameter \cite{Macdonald-5, Ngai}. So,
according to the present analysis $\beta_{KWW}=\alpha\beta$, where $\alpha$ and
$\beta$ represent the H--N shape parameters of the ionic polarization mechanism in $\varepsilon^*$ formalism. The
connection between the H--N and KWW parameters appears also in other relevant works in the literature
\cite{Havriliak,Alvarez, Boese}.

\section{Conclusions}

Summarizing this work, the BNN relation has been extracted while it has 
been found that $p$ coefficient depends on the H--N shape parameters $\alpha$ and $\beta$ 
as well as on the ratio $\omega_0/\omega_{HN}$ of two frequencies which are directly 
connected to characteristic time scales of ions motion (Eq.(\ref{bnn-13})). This equation is compared 
to the BNN relation (Eq.(\ref{bnn-16})) which could be extracted based on Dygas findings in 
his previous work. Both approaches highlight the dependence of BNN 
coefficient from the same parameters. In C--D dielectric behavior of ionic 
polarization mechanism, both relationships are nearly identical 
predictions with very high accuracy for the $p$ coefficient, while 
deviations are observed in the C--C one.
Representative simulations have shown that only Eq.(\ref{bnn-16}), is consistent 
with the power law frequency dependence of AC conductivity in symmetrical C--C cases. The 
identification of $p$ values ​​in asymmetric C--D behavior indicates that 
in these cases, the function $\sigma'(\omega)$ should be approached very well 
from the power law not only at high frequency limit but also at $\omega=\omega_0$.
Both expressions of BNN relation could result in many possible 
$p$ values. However, the response of matter could exhibit individual 
characteristics. In both cases, of C--D and C--C behavior, the $p$ 
values, depending on $\omega_0/\omega_{HN}$ ratio, lie within the range where 
the majority of the $p$ values ​​recorded in the literature falls. 
Specifically, $p$ values ​​close to unity arise for the lower values 
​​of the ratio $\omega_0/\omega_{HN}$ and for an appreciable range of the H--N 
shape parameters $\beta$ and $\alpha$.

The two expressions of BNN coefficient, applied to experimental 
data of a crosslinked polyurethane, which satisfies the BNN relation
and the ionic polarization mechanism exhibits C--D behavior. Both predictions 
are in excellent agreements with the values ​​calculated from 
original BNN relation, according to experimental data analysis. The 
results also showed a trend of gradual reduction of the $p$ coefficient, as 
the temperature increases.
 
\noindent
{\bf Acknowledgements} 

We would like to express our gratitude to the referee for his helpful and motivational comments which have enable us 
to clarify many points in our work.

\end{document}